\documentclass[a4paper,11pt]{article}
\usepackage{pos}
\usepackage{dsfont}

\title{Vector-like Quarks}

\author*[a]{Gustavo Castelo Branco}
\author[a]{M. N. Rebelo}

\affiliation[a]{Centro de F\'isica Te\'orica de Part\'iculas, CFTP, Departamento de F\'\i sica,\\ Instituto Superior T\'ecnico, Universidade de Lisboa,\\
Avenida Rovisco Pais, 1049-001 Lisboa, Portugal}

\emailAdd{gbranco@tecnico.ulisboa.pt}
\emailAdd{rebelo@tecnico.ulisboa.pt}

\abstract{
In this talk we emphasise  the importance of vector-like quarks (VLQs) and their potential to solve some of the open questions of the Standard Model. These are, in some sense minimal extensions of the Standard Model, that can be probed in the next round of experiments.
We also make an analogy between vector-like quarks(VLQs) and right-handed neutrinos, emphasising  that in both cases some of the flavour dogmas of the SM are violated in a controlled way.
}

\FullConference{%
  7th Symposium on Prospects in the Physics of Discrete Symmetries (DISCRETE 2020-2021)\\
  29th November - 3rd December 2021\\
 Bergen, Norway}

\begin{document}
\maketitle

\section{Introduction}

The Standard Model (SM) is about 50 years old! One may consider that the SM was finally born in 1971, with the discovery of renormalizability of spontaneously broken gauge theories by 't Hooft and Veltman \cite{tHooft:1971akt,tHooft:1971qjg,tHooft:1972tcz}. The birth took about ten years (1961-1971). The building of the SM started in 1961 with Glashow's paper \cite{Glashow:1961tr} suggesting the gauge group $ SU(2)\times U(1)$. One may ask the question why to include 1971 as the year when the construction of the SM was completed. The best way to answer this question is by asking the following question: What were particle physicists doing in 1968? One may answer this question consulting the ICHEP(1968) Proceedings \cite{Prentki:1968fha}. This was one of the most 
exceptional conferences 
of all times. The number of Nobel laureates (either already Nobel at the time of the conference or later) is truly impressive: 19! But Nobody among these renowned participants talked about gauge theories! The turning point was induced by 't Hooft and Veltman's revolutionary papers. This illustrates the importance of the work of 't Hooft and Veltman in completing the construction of the SM. \\

Strictly speaking, the SM as suggested by Glashow, Weinberg and Salam \cite{Glashow:1961tr,Weinberg:1967tq,Salam:1968rm}, has been ruled out by the discovery of neutrino oscillations indicating that at least two neutrinos have non-zero masses. However, most of the predictions of the SM involving the gauge sector, the quark sector and the Higgs sector have been verified experimentally to a great precision, with only some experimental hints for physics beyond the SM in these sectors, yet to be confirmed. Furthermore, there is indirect evidence for the need of new sources of CP violation in order to account for the observed baryon asymmetry of the Universe (BAU). The SM also lacks suitable dark matter candidates and leaves several questions open, in particular it does not provide an understanding of the flavour pattern of fermion masses and mixing. \\

At this stage, one may ask: What is the key message that ``Nature" is telling us concerning the observation of neutrino oscillations? In our opinion the message is: Choose Simplicity! Indeed the SM is ``almost" the simplest possibility to put together the charged  and neutral gauge interactions including the electromagnetism. Why almost? The reason for this adjective has to do with the fact the new SM (NSM) is simpler. By NSM we mean the SM with the addition of right handed neutrinos. Recall that at the time one of the rules to construct gauge models was: Write down the most general Lagrangian consistent with gauge invariance and renormalizability. If one follows this rule one has to include Dirac mass terms for the neutrinos, $\overline{\nu^0_L} m \nu^0_R$ together with Majorana mass terms, $\frac{1}{2} M\nu^{0T}_R C^* \nu^0_R$. This automatically leads to the seesaw mechanism 
\cite{Minkowski:1977sc,Yanagida:1979as,Glashow:1979nm,GellMann:1980vs,Mohapatra:1979ia}.
Peter Minkowski was the first one to write the seesaw formula \cite{Minkowski:1977sc} but had forgotten until hearing about it again in 2004. Historically there were other motivations to introduce right handed neutrinos: SO(10) grand unified theories (GUTs) and  Left-Right Symmetric models, for a review see \cite{Mohapatra:2006gs} and references therein. At the moment it is not known whether neutrinos are Dirac or Majorana particles. But if we follow the ``Simplicity Principle" neutrinos should be Majorana particles.  For a long time, there was a profound prejudice in favour of massless neutrinos based on the fact that there was no experimental evidence for neutrino masses and leptonic mixing. This prejudice was extended to GUTS. The only ``reasonable" GUT was SU(5). In SU(5) right handed neutrinos are not introduced and B-L is an accidental exact symmetry. As a result, neutrinos are strictly massless in SU(5), just as  in the SM. \\

There were other related dogmas in Flavour Physics, based on the properties of the SM and on stringent experimental limits, namely:
\begin{itemize}
\item
{Dogma 1 : There should be no Z-Mediated Flavour-Changing Neutral Currents (FCNC) at tree level either in the quark or in the leptonic sectors. }
\item{Dogma 2: Dogma 1 was extended to the scalar sector. There should be no scalar mediated FCNC at tree level.}
\end{itemize}

Dogma 1 is violated in the leptonic sector with Majorana neutrinos.
Dogma 2 is safely violated in Branco-Grimus-Lavoura (BGL) models \cite{Branco:1996bq} with an extended Higgs sector and their generalisations by Botella, Branco and Rebelo \cite{Botella:2009pq}.  Here safely refers to the natural suppression of these effects, based on a symmetry.\\

In the seesaw framework, with the addition of three right handed neutrinos,  the leptonic mixing matrix at low energies, the Pontecorvo-Maki-Nakagawa-Sakata, $U_{PMNS}$, matrix, is not $3\times 3$ unitary. By the same token Dogma 1 is also violated. These two aspects result from the mixing among the six neutrino fields appearing in the flavour basis and leading to the three light neutrinos and the three heavy neutrinos. This is a special feature of seesaw since Majorana neutrinos have fewer degrees of freedom than Dirac neutrinos, resulting in the possibility of doubling the number of physical particles with different masses.  Indeed FCNC appear at tree level but are naturally suppressed by the ratio m/M where m stands for Dirac neutrino masses and M for the Majorana masses of right handed neutrinos. Since the Majorana right handed neutrino mass terms are gauge invariant they are  not protected by the gauge symmetry, unlike Dirac mass terms, and thus can be much larger than v, the scale of electroweak symmetry breaking. The same suppression operates in the deviations of unitarity of the  $U_{PMNS}$ matrix.

\section{Analogy between Vector-like Quarks (VLQs) and right-handed neutrinos}
Consider an extension of the SM where new quark isosinglets, i.e., singlets of SU(2), of charge -1/3 and/or  +2/3 are introduced, such that the right handed and the left handed components transform in the same way under the gauge group. In this case mass terms of the form $M_D \overline{D_L}D_R$ and/or $M_U \overline{U_L}{U_R}$  are gauge invariant and must be introduced in the Lagrangian. The simplest possibility is to assume that VLQs are isosinglets but there are other possible choices of representations that are also of interest. VLQs have in common with right handed neutrinos the fact that a new scale (let us call it V) is introduced by these mass terms. This scale can be significantly above v, the scale of electroweak symmetry breaking. Like in the leptonic sector, 
VLQs lead to violations of $3 \times 3$ unitarity of  the Cabibbo-Kobayashi-Maskawa matrix, 
the $V_{CKM}$ matrix, as well as to Z-mediated FCNC. Both effects are related. In the case of vector-like quarks these effects result from the mixing of the standard-like quarks with the new vector-like quarks and are naturally suppressed by the ratio v/V. In the quark sector all physical particles are of Dirac-type.  Majorana mass terms would break any U(1) charge and therefore are forbidden for any particle with non-zero electric charge.


\section{What can Vector-like Quarks (VLQs) do for you?}
Next we give a partial answer to the above question, providing a list of some of the
contributions that VLQs can give to an understanding of the some of the open questions
in the SM. \\

i) VLQs provide a simple alternative solution to the Strong CP problem \cite{tHooft:1976rip,tHooft:1976snw},
without having axions, which so far have not been seen. This solution was proposed 
by Barr and Nelson \cite{Nelson:1983zb,Nelson:1984hg,Barr:1984qx}
and a minimal implementation in a model, was given by Bento, Branco and 
Parada \cite{Bento:1991ez}. \\

ii) They provide the simplest extension of the SM with spontaneous CP violation, in a model consistent with experiment. At this stage it is worth recalling what are the requirements 
in order to have a viable model of spontaneous CP violation:  The Lagrangian should be CP 
invariant but CP invariance should be broken by the vacuum. The minimal extension of the SM leading to spontaneous CP violation involves the extension of the scalar sector through the addition of a complex singlet. Sometimes the vacuum leads to geometrical vacuum phases. It is important to note that often these geometrical phases do not lead to spontaneous CP violation \cite{Branco:1983tn}.

The vacuum phase should be able to generate a complex Cabibbo-Kobayashi-Maskawa matrix, $V_{CKM}$, since experimentally we do know that $V_{CKM}$, is complex even if one allows for the presence of New Physics \cite{Botella:2005fc,UTfit:2005lis}.
Note that at present we do know that the phase $\gamma$ where 
$\gamma = \arg (- V_{ud}V_{cd}V^*_{ub}V^*_{cd})$
is not zero or pi and we also know that it cannot be generated solely through new physics beyond the SM.\\

iii) With the additional introduction of right handed neutrinos in the leptonic sector it is possible to build models with VLQs where there is a common origin for all CP violations \cite{Branco:2003rt}, 
i.e., models providing  at the same time a solution for the strong CP problem of the type described in i), spontaneous CP violation, CP violation in the quark sector in agreement with experiment of the type described in ii) and CP violation in the leptonic sector both at low energies and at hight energies leading to the possibility of  leptogenesis \cite{Fukugita:1986hr}. Such models require minimal extensions of the SM with either the introduction of one down vector-like or one up vector-like quark and will be described in what follows. \\

iv) VLQs provide a simple framework where there are New Physics contributions to 
$B_d-{\overline B_d}$, $B_s-{\overline B_s}$ and/or $D^0-{\overline D^0}$ mixing.
Also in models with up-type VLQs, there are  contributions to the decay 
$t \rightarrow  c Z_\mu$ at tree level \cite{Botella:2012ju}. \\

v) VLQs may populate the ``desert" between v and some higher mass scale like $M_{GUT}$, without
worsening the hierarchy problem. To our knowledge, this was first emphasised in a paper by Pierre Ramond \cite{Ramond:1981jx}. \\

vi) VLQs may contribute to the unification of the coupling constants without requiring Supersymmetry \cite{Dermisek:2012ke}. \\

vii) VLQs may play an important r\^ ole in providing an explanation for the $V_{CKM}$, unitarity problem, arising from new measurements of $|V_{us}|$ and $|V_{ud}|$  together with new theory calculations and lattice results \cite{Seng:2018yzq, Seng:2018qru,Czarnecki:2019mwq,Seng:2020wjq,Hayen:2020cxh,Shiells:2020fqp,Aoki:2021kgd}, namely:
\begin{equation}
\Delta \equiv 1 - |V_{ud}|^2 - |V_{us}|^2 - |V_{ub}|^2 < 1, \qquad \sqrt{\Delta } \sim 0.04                         
\end{equation}
at the level of 2 to 3 standard deviations \cite{Belfatto:2019swo,Belfatto:2021jhf,Branco:2021vhs,Botella:2021uxz}. 
One may wonder whether or not this two to three sigma deviation  should be taken as a serious hint for physics beyond the SM. Several experts are taking it seriously.

\section{A Common Origin for all CP Violations}
Let us illustrate how VLQs can address the first three open questions listed above \cite{Branco:2003rt}. The model is based on the Bento, Branco and  Parada framework  \cite{Bento:1991ez} together with the leptonic sector of the SM extended with the inclusion of three right handed neutrinos. CP invariance is imposed on the Lagrangian so that all coefficients are taken to be real. A $\mathds{Z}_2$ symmetry is imposed to the quark and to the Higgs sectors enlarged with the addition of one down-type isosinglet vector-like quark ($D^0_L, D^0_R$)  and a complex scalar singlet $S$. 
\begin{equation}
\left( 
\begin{array}{c}
u^0 \\ 
 d^0
\end{array}
\right)_{iL}, u^0_{iR},  \ \ d^0_{\alpha R}, \ \ D^0_L, \ \ i= 1,2,3, 
\ \   \alpha = 1, ..., 4, \ \ \phi, \ \  S
\end{equation}
All standard like particles in these two sectors are invariant under this $\mathds{Z}_2$ symmetry while $D^0_L, D^0_R$ and $S$ are odd under $\mathds{Z}_2$.
The $SU(2) \times U(1) \times \mathds{Z}_2$ invariant potential is of the form:
\begin{equation}
V = V_0 \ (\phi, S) + (\mu^2 + \lambda_1 \ S^\ast S + \lambda_2 \ \phi^\dagger \phi) (S^2 + S^{\ast 2}) + \lambda_3 \ (S^4 + S^{\ast 4} )
\end{equation}
where $V_0$  contains all terms that are phase independent and includes the SM Higgs potential. The vevs of the neutral scalar fields will be of the form:
\begin{equation}
\langle {\phi}^0 \rangle = \frac{v}{\sqrt 2}, \ \   \ \ \   
\langle S \rangle = \frac{V \exp (i \beta )}{\sqrt 2}
\label{eq:vevphiS}
\end{equation}
and will in general violate CP. There are two mass scales $v$, the electroweak breaking scale and $V$ which can be much larger than $v$. The Yukawa interactions of the quarks are given by:\begin{equation}
\label{BBP:yukawa} 
 {\cal L}_Y = - \sqrt{2} (\overline{u^0} \ \overline{d^0})_L^i (g_{ij} \phi \, d^0_{jR} + 
h_{ij}  \tilde{\phi}\, u^0_{jR}) - M_d \overline{D^0_L}\, D^0_R - \sqrt{2} (f_i \, S + f^\prime_i \, S^* ) \, \overline{D^0_L} d^0_{iR} + \mbox{h.c.}
\end{equation}
with (i,j = 1, 2, 3) and the down quark mass matrix is now of the form:
\begin{equation}
{ \cal M}_d = \left( \begin{array}{cc}
m_d & 0 \\
\overline{M_d} & M_d
\end{array} \right)
\end{equation}
the zero block of ${ \cal M}_d$ results from the $\mathds{Z}_2$ symmetry. Only the block $\overline{M_d}$ can be complex. Therefore $\arg \det {\cal M}_d$ is zero.  The up quark mass matrix is real. As a result 
$\arg (\det {\cal M}_d \times \det m_u)$ is zero and the strong CP problem is solved \` a la Barr and Nelson \cite{Bento:1991ez}.
Mixing of the four down type quarks can lead to a complex $V_{CKM}$ with deviations from unitarity suppressed by the ratio $v/M_D$, where $M_D$ is the mass of the heavy down quark,  as was shown in \cite{Bento:1991ez}. ZFCNC also feel the same type of suppression \cite{Bento:1991ez}. However the strength of CP violation does not suffer the same type of suppression provided that the new mass scale of $M_d$ is not higher than that of $\overline{M_d}$ \cite{Bento:1991ez}, even in the limit when both go to infinity. This is an example of non-decoupling of very massive particles.  In this framework the introduction of the complex scalar singlet $S$ together with the Higgs doublet 
leads to spontaneous CP violation. The CP violating phase $\beta$ generated at high energies leads to  a complex $V_{CKM}$  matrix due to the presence of the isosinglet vector-like quark which has couplings  to the scalar $S$ together with  SM-like right handed quarks. This r\^ ole can be played by either a down or an up-type vector-like quark.\\

In the extension of this model to the leptonic sector the $\mathds{Z}_2$ is promoted to a 
$\mathds{Z}_4$ symmetry \cite{Branco:2003rt} under which the fields that are not invariant transform as:
\begin{equation}
\begin{aligned}
D^0 &\rightarrow -D^0, \quad S \rightarrow -S\,,   \\
{\psi _ L^0} &\rightarrow i \psi _ L^0, \quad
e_R^0 \rightarrow i e_R^0,\quad   
\nu_{R}^0 \rightarrow i \nu_{R}^0\,,
\label{eq:zds}
\end{aligned}
\end{equation}
where $\psi^0_L$ are the left handed lepton doublets, $e^0_R$ are the right handed charged leptons and $\nu^0_R$ stand for the three right handed neutrinos. The Yukawa terms for the leptonic sector are:
\begin{equation}
{\cal L}_l = {\overline {\psi _ L^0}} G_l \phi \ e_R^0 +
{\overline {\psi _ L^0}} G_{\nu}  \tilde{\phi } \ \nu _R^0 +
\frac{1}{2} {\nu} _R^{0T} C ({f_\nu} S + 
+ {f_\nu}^{\prime} S^\ast )\nu _R^0 +  h. c.   
\label{eq:ll}
\end{equation} 
again all new coefficients are taken to be real.
After spontaneous symmetry breakdown one obtains the following mass terms for the charged leptons and for the neutrinos:
\begin{equation}
\begin{aligned}
{\cal L}_m  &= -\left[ \overline{{\nu}_{L}^0} m \nu_{R}^0 +
\frac{1}{2} \nu_{R}^{0T} C M \nu_{R}^0+
\overline{l_L^0} m_l l_R^0 \right] + 
{\rm h. c.}  \\
&= - \left[ \frac{1}{2} n_{L}^{T} C {\cal M}^* n_L +
\overline{l_L^0} m_l l_R^0 \right] + {\rm h. c.}\,.
\label{eq:lm}
\end{aligned}
\end{equation}
The leptonic mass matrices are now given by:
\begin{equation}
\begin{aligned}
{\cal M} &= \left(\begin{array}{cc}
0 & m \\
m^T & M \end{array}\right), \ \ m_l = \frac{v}{\sqrt 2} G_l, 
\ \  m =  \frac{v}{\sqrt 2} G_{\nu} \\
M &=  \frac{V}{\sqrt 2}( f_{\nu}^+ \cos (\alpha) +
i f_{\nu}^- \sin (\alpha) ) 
\label{eq:mmm}
\end{aligned}
\end{equation}
with $f_{\pm}^{\nu} \equiv f_{\nu}  \pm  
{f_{\nu} }^{\prime}$. 
In Ref.~\cite{Branco:2003rt} it is shown how this framework can lead to a complex $U_{PMNS}$ matrix.  It is also shown that this framework contains the necessary ingredients allowing for the possibility of leptogenesis \cite{Fukugita:1986hr}. Although CP violation in the leptonic sector has not yet been observed it is natural to expect that it should also be present in this sector since 
it is broken in the quark sector and therefore it is not a symmetry of nature.

\section{Conclusions}

We have emphasised that vector-like quarks (VLQs) are one of the simplest extensions of the SM and yet they provide, on the one hand a possible solution to some of the open questions of the SM
and on the other hand, they predict various new phenomena so that their predictions may be probed in the next round of experiments. Interesting aspects of VLQs include the possibility of having a realistic model with spontaneous CP violation and one of the simplest solutions to the strong CP problem without the introduction of axions. VLQS also provide a framework where one can have a common origin for all manifestations of CP violation, including CP breaking in the quark and lepton sectors and the generation of the Baryon Asymmetry  of the Universe through leptogenesis. \\

In section 3 we have summarised several aspects of physics beyond the Standard Model that can be addressed via the introduction of VLQs. Many of these are related to recent experimental results from collider experiments. This list is far from complete. \\

Recent measurements of $|V_{us}|$ and $|V_{ud}|$, together with recent theoretical improvements in calculations, indicate that unitarity of the first row of the $V_{CKM}$ may be violated at the level of two or three standard deviations. It is possible to account for these deviations with simple extensions of the Standard Model with vector-like quarks. This possibility was analysed in detail in \cite{Belfatto:2019swo,Belfatto:2021jhf,Branco:2021vhs,Botella:2021uxz}. \\

We have  addressed in different works of ours, involving VLQs, other recent experimental hints for new physics. However, we have opted to illustrate in section 4 of this paper some of the mechanisms at work in models with VLQs rather than to deal with very specific phenomena. \\

Models with VLQs have always attracted attention in the literature. At present many more authors are turning their attention to the study of this class of models. VLQs naturally arise  in the context of some Grand Unified models like $E_6$.

\acknowledgments
The authors thank the local organisers of the symposium Discrete 2020-2021 for the very fruitful scientific meeting and the warm hospitality. G.C.B. thanks Bill Marciano for sharing with us his views about the significance of the CKM unitarity problem. This work was partially supported by Funda\c c\~ ao para a Ci\^ encia e a Tecnologia (FCT, Portugal) through the projects CFTP-FCT Unit UIDB/00777/2020 and UIDP/00777/2020, PTDC/FIS-PAR/29436/2017, CERN/FIS-PAR/0008/2019 and CERN/FIS-PAR/0002/2021 
which are partially funded through POCTI (FEDER), COMPETE, QREN and EU.

\bibliographystyle{JHEP}


\end{document}